\documentclass[prd,onecolumn,floatfix,letterpaper,amsmath,amssymb,nofootinbib,superscriptaddress]{revtex4}
\usepackage{graphicx}
\usepackage{epsfig}
\usepackage{bm}
\usepackage{hyperref}
\usepackage{color} 
\usepackage{amsmath}
\usepackage{amssymb}
\usepackage{amsthm}
\usepackage{amsfonts}
\usepackage{latexsym,float,url}
\begin{document}

   \title{Causal structures in Gauss-Bonnet gravity}

\author{Keisuke Izumi}
 \email{izumi@phys.ntu.edu.tw}
 \affiliation{Leung Center for Cosmology and Particle Astrophysics,
 National Taiwan University, Taipei 10617, Taiwan}

\date{\today}

\begin{abstract}

We analyze causal structures in Gauss-Bonnet gravity. 
It is known that Gauss-Bonnet gravity potentially has superluminal propagation of gravitons due to its noncanonical kinetic terms. 
In a theory with superluminal modes, an analysis of causality based on null curves makes no sense, and thus, we need to analyze them in a different way. 
In this paper, using the method of the characteristics, we analyze the causal structure in Gauss-Bonnet gravity. 
We have the result that, on a Killing horizon, gravitons can propagate in the null direction tangent to the Killing horizon. 
Therefore, a Killing horizon can be a causal edge as in the case of general relativity, i.e. 
a Killing horizon is the ``event horizon" in the sense of causality. 
We also analyze causal structures on nonstationary solutions with $(D-2)$-dimensional maximal symmetry, including spherically symmetric and flat spaces. 
If the geometrical null energy condition, $R_{AB}N^AN^B \ge 0$ for any null vector $N^A$, is satisfied, the radial velocity of gravitons must be less than or equal to that of light. 
However, if the geometrical null energy condition is violated, gravitons can propagate faster than light. 
Hence, on an evaporating black hole where the geometrical null energy condition is expected not to hold, classical gravitons can escape from the ``black hole" defined with null curves.
That is, the causal structures become nontrivial. 
It may be one of the possible solutions for the information loss paradox of evaporating black holes.

\end{abstract}

\maketitle
\section{Introduction}

Quantum gravity is one of ultimate goals in fundamental physics. 
Many models of quantum gravity have been proposed.
Some of them lead to an effective theory with noncanonical kinetic terms in the low-energy limit. 
In such theories the maximum speeds are different for different fields~\cite{Gibbons:2000xe} and potentially  superluminal modes appear~\cite{Aragone,ChoquetBruhat:1988dw,Brigante:2007nu,Brigante:2008gz}. 
Meanwhile, to solve the so-called dark energy and dark matter problems, various theories of modified gravity have been proposed. 
Some of them also involve superluminal propagations~\cite{Chiang:2012vh,Izumi:2012qj,Ong:2013qja,Izumi:2013poa,Deser:2013eua,Izumi:2013dca,Deser:2013qza}.

In general relativity with fields having canonical kinetic terms, 
the speeds of all modes are less than or equal to that of light, and then 
we analyze causal structures based on null curves. 
This is justified by the fact that any modes cannot go through null hypersurfaces in a spacelike direction. 
However, if a theory has superluminal modes, i.e. spacelike propagations, 
the discussion based on null curves makes no sense. 
We must analyze causal structures with the fastest propagations. 
This is essential, for instance, in the definition of black holes. 
Usually, we define a black hole as the outside of the chronological past of the future time infinity.
Here, the chronological past is defined with null curves.  
In contrast, if we have superluminal modes, the chronological past defined with null curves does not show the causal structures and
we need to define the ``chronological past" in the sense of causality  with the fastest modes. 
This cannot be analyzed only with the metric. 
Information regarding the propagations is needed. 

With superluminal propagations, the information loss paradox of evaporating black holes may be solved. 
Superluminal propagations can convey the information from inside of the black hole to the outside. 
Evaporating black holes are semiclassical objects. 
There, we consider the quantum effects of matter fields on classical geometry. 
Namely, we must deal with matter fields as quantum objects, while gravity is classical. 
Therefore, for the causal analysis of gravity, we can use classical physics, 
which is much easier than the discussion of quantum causality for matter fields. 
We expect that the property of the causal structure is similar even for quantum matter fields. 
In this paper, as a first step in the analysis of causal structures on evaporating black holes, we deal with the easiest modes, that is, gravitons.
As a lowest-order correction of gravity theory, we consider the Gauss-Bonnet correction term.

Gauss-Bonnet gravity is a natural extension of  general relativity in higher dimensional spacetime. 
In spite of the fact that the action has the curvature-squared terms, 
the equation of motion for gravity has up to the second-order derivatives of metric~\cite{Boulware:1985wk,Zumino:1985dp,Cho:2001su},
which prevents the theory from ghost excitations. 
Moreover, the theory is interesting  because it is realized in the low-energy limit of heterotic string theory~\cite{Gross:1986iv,Gross:1986mw,Metsaev:1987bc,Zwiebach:1985uq,Metsaev:1987zx}. 
Gauss-Bonnet gravity is studied in many contents, such as black holes~\cite{Dotti:2007az,Barrau:2003tk,Torii:2005xu,Charmousis:2008kc,Cai:2003gr,Cho:2002hq,Cai:2001dz}, braneworld model~\cite{Kim:1999dq,Cho:2001nf,deRham:2006pe,Kofinas:2003rz,Brown:2006mh,Charmousis:2002rc,Maeda:2003vq,Maeda:2007cb,BouhmadiLopez:2012uf,Bouhmadi-Lopez:2013gqa,Yamashita:2014cra,Liu:2014xja}, AdS/CFT correspondence~\cite{Brigante:2007nu,Brigante:2008gz,Buchel:2008vz,Gregory:2009fj,Buchel:2009sk,Hung:2011xb,Buchel:2009tt,Camanho:2009vw} and so on~\cite{Nozawa:2007vq,Maeda:2007uu,Maeda:2011ii,Maeda:2008nz,Charmousis:2008ce}.

It is well known that Gauss-Bonnet gravity theory involves superluminal propagation of gravitons; this was noted in early works~\cite{Aragone,ChoquetBruhat:1988dw} and also in recent works in the context of the AdS/CFT correspondence~\cite{Brigante:2007nu,Brigante:2008gz}.
However, the concrete analysis on general manifolds has not been done. 
The purpose of this paper is that, with less assumptions, we show generic properties of causal structures. 
We basically consider two cases: 
one is the locally stationary spacetime, and the other is spacetime with $(D-2)$-dimensional maximal symmetry.

The organization of this paper is as follows.
In Sec.~\ref{SuperLum}, we show the origin of superluminality with an example of a scalar field. 
We also explain the relation between superluminality and acausality.
In Sec.~\ref{review}, we briefly review the method of characteristics.
In Sec.~\ref{GB}, we define Gauss-Bonnet gravity which we analyze.
In Sec.~\ref{cha}, we derive the characteristic equations of Gauss-Bonnet gravity. 
We give the contributions stemming from the Einstein-Hilbert and the Gauss-Bonnet terms in 
Sec.~\ref{gene} and in Sec.~\ref{gb}, respectively. 
In Sec.~\ref{gene}, we also show that, in  general relativity, the characteristic hypersurface for gravitons always becomes null.
In Sec.~\ref{LS}, we analyze the causal structures in stationary cases, 
while in Sec.~\ref{DB} we consider cases with $(D-2)$-dimensional maximal symmetry.
Finally, we summarize our work with a discussion in Sec.~\ref{Sum}.

We use the following notation for indices.
Large Latin letters $\{A,B,\ldots \}$ are the indices for the $D$-dimensional spacetime, while  Greek letters $\{\mu,\nu,\ldots \}$ are the indices for the $(D-1)$-dimensional hypersurface $\Sigma$ that we concentrate on. 
The index ``$0$" means the direction which is \emph{not} tangent to $\Sigma$.
We use the index ``$1$" for the null direction on the hypersurface $\Sigma$ if it is null, or 
in Sec.~\ref{DB} for the direction which is normal to Killing directions on the hypersurface $\Sigma$ (roughly speaking, the radial direction in the spherically symmetric case).
We denote the normal directions to the $0$ and $1$ directions by the small Latin letters $\{i,j,\ldots \}$.

\section{Theory with Superluminal Modes}\label{SuperLum}

In the standard theory, causal structures are discussed with null curves. 
Here, ``the standard theory" means that in the theory all the fields have canonical kinetic terms. 
In such a theory, the highest speeds are the same as that of light, which propagates in  null direction. 
The causally related region, i.e. the Cauchy development, is configured with the fastest propagation, 
and thus, we can justify the causal structures based on null curves. 
However, if a theory has superluminal modes, the situation becomes different. 
We must analyze causal structures based on the fastest propagations.

Now, the question is: which theory has superluminal modes. 
One example is Gauss-Bonnet gravity, where the propagations of gravitons can be superluminal on a nontrivial background. 
This was pointed out at the end of the 1980s~~\cite{Aragone,ChoquetBruhat:1988dw} and recently discussed in the AdS/CFT context~\cite{Brigante:2007nu,Brigante:2008gz}. 

We show the reason why superluminal modes appear by using a scalar field example. 
We first consider a scalar field theory with a canonical kinetic term, whose equation of motion is written as 
\begin{eqnarray}
g^{AB} \nabla_A\nabla_B \phi + V(\phi)=0.
\end{eqnarray}
``Canonical kinetic term" means the coefficient of the kinetic term (i.e. the second-order derivative term) is proportional to the metric $g^{\mu\nu}$. 
To see the maximum speed of a propagation for $\phi$, we take the high-energy limit, 
where we can ignore the potential term. 
In the Fourier space, the equation becomes 
\begin{eqnarray}
g^{AB} k_A k_B \phi_k =0,
\end{eqnarray}
where $\phi_k$ is a Fourier mode of $\phi$ with momentum $k_{A}$.
This gives the solution that $k_A$ is null. 
However, if a theory has a noncanonical kinetic term, the situation changes. 
For instance, we consider a scalar field $\tilde \phi$ with the following equation:
\begin{eqnarray}
\left(g^{AB} + \alpha\nabla^A\psi \nabla^B\psi \right)\nabla_A\nabla_B \tilde \phi + V(\tilde \phi)=0,
\end{eqnarray}
where $\psi$ is another scalar field and $\alpha$ is a constant. 
The kinetic term has the coefficient $\left(g^{AB} + \alpha\nabla^A\psi \nabla^B\psi \right)$. 
Taking the high frequency limit for $\tilde \phi$, we can again neglect the potential term and in the Fourier space for $\tilde \phi$ we have
\begin{eqnarray}
\left(g^{AB} + \alpha\nabla^A\psi \nabla^B\psi \right) \tilde k_A \tilde k_B \tilde \phi_{\tilde k} =0. \label{scaexa}
\end{eqnarray}
Then, $\tilde k_A$ is a null direction for the effective metric $\left(g^{AB} + \alpha\nabla^A\psi \nabla^B\psi \right)$, which is different from that for the real metric $g^{AB}$, i.e. $k_A$. 
Therefore, with nonzero $\nabla_A \psi$ the fastest mode does not follow a null trajectory and 
can be spacelike or timelike depending on the value of $\alpha$ and $\nabla_A \psi$. 
With some values of $\alpha$ and $\nabla_A \psi$, the effective metric can be Euclidean, or from a Euclidean metric we can construct a Lorentzian effective metric~\cite{Mukohyama:2013ew,Mukohyama:2013gra,Kehayias:2014uta}. 

Gravity theory has generically nonlinear kinetic terms, and thus, it is not trivial that the maximum speed of gravitons becomes the same as that of light. 
Although the kinetic term looks complicated in general relativity, we can check that the kinetic term for gravitons takes the canonical form. 
(See Sec.~\ref{gene}.)
However, if we consider an extension of general relativity, it can easily break the canonical structure of the kinetic terms.
Since the Gauss-Bonnet term directly gives second-order derivative terms in the equation of motion, adding it results in a nontrivial form of the kinetic terms. 
Moreover, because general relativity is a system with constraints, even if the added terms 
do not have derivatives, it leads to a modification of the structures in the kinetic terms after solving the constraint equations.
In massive gravity, for instance, even though the modification of general relativity is just adding mass terms, i.e. no derivative terms, superluminal modes appear~\cite{Chiang:2012vh,Izumi:2013poa,Deser:2013eua,Deser:2013qza}.

We stress that superluminality does not directly result in acausality, which means the existence of bad causal structures such as a closed curve of propagations. 
If both a theory and a state have Lorentz symmetry, superluminality causes acausality. 
If a superluminality mode exists, due to  Lorentz symmetry it can be adjusted to any spacelike direction and we can easily construct closed curves. 
However, if  a state does not have  Lorentz symmetry, we cannot use this discussion.
In the above example for a scalar field (\ref{scaexa}), only when $\nabla^A \psi$ has a nonzero value, a nontrivial propagation appears. 
Then, nonzero $\nabla^A \psi$ breaks the Lorentz symmetry. 
Similarly, in Gauss-Bonnet gravity, only if the curvature is not zero, which breaks Lorentz symmetry, the propagations of gravitons become nontrivial.
In such theories, we need to check the causal structure of each solution.

\section{Brief Review of Characteristics}\label{review}

We briefly review the method of characteristics, which is a powerful tool for analyzing causal structures~\cite{Izumi:2013poa,Deser:2013eua,Deser:2013qza,Ong:2013qja,Izumi:2013dca,CH,Nester8}. 
The method shows the hypersurface beyond which the evolution equations cease to give an unique solution. 
This is mathematically characterized as the hypersurface where the coefficients of the highest-order derivative with respect to its nontangent direction vanish. 
This can be intuitively understood as follows. 
To solve a $N$th order differential equation, generically it is only necessary to impose $N$ initial conditions for up to $(N-1)$-th order derivatives. 
Namely, the evolution of $i$th order derivatives with $0<i<N-2$ is uniquely fixed by the given initial condition for $(i+1)$-th order derivatives, 
while the evolution for the $(N-1)$-th order derivative, that is $N$th order derivative, is obtained from the equation. 
However, if the coefficient of the $N$th order derivative vanishes, we can never solve the equation for the $N$th order derivative, and thus, the evolution of the $(N-1)$-th order derivative cannot be fixed.

Let us see the details in the case of a partial differential equation. 
We derive the hypersurface, denoted by $\Sigma$, beyond which the evolution is not unique. 
Such a hypersurface is called the \emph{characteristic hypersurface}. 
We define a vector $\xi^A$ which is \emph{not} tangent to $\Sigma$. 
(Usually, $\xi^A$ is chosen to be normal to $\Sigma$ for simplicity. However, it makes a null limit complicated. Therefore, in this paper, we do \emph{not restrict} $\xi^A$ to be the normal vector to $\Sigma$.) 
Suppose we have a quasilinear equation for a variable $\phi$,\footnote{
$\phi$ does not need to be a scalar field.
}
\begin{eqnarray}
M^{A_1,\cdots,A_N} \partial_{A_1} \cdots \partial_{A_N} \phi + {\cal O}\left(\partial^{N-1} \phi\right) =0.
\end{eqnarray}
Here, a \emph{quasilinear equation} means the highest-order derivative appears linearly. 
This is the necessary condition for an unique evolution.\footnote{
Exactly stated, the necessary condition is that the equation is linear for $\xi^{A_1}\cdots\xi^{A_N}\partial_{A_1} \cdots \partial_{A_N} \phi$.
}
We decompose the equation  along  the lines of the Arnowitt-Deser-Misner formalism~\cite{ADM1,ADM2} with the understanding that $\xi^A$ may be nontimelike. 
Then, the condition of the characteristic hypersurface is that the coefficient of $\xi^{A_1}\ldots \xi^{A_N}\partial_{A_1} \ldots  \partial_{A_N} \phi$ becomes zero.

\begin{figure}[tbp]
  \begin{center}
    \includegraphics[keepaspectratio=true,height=50mm]{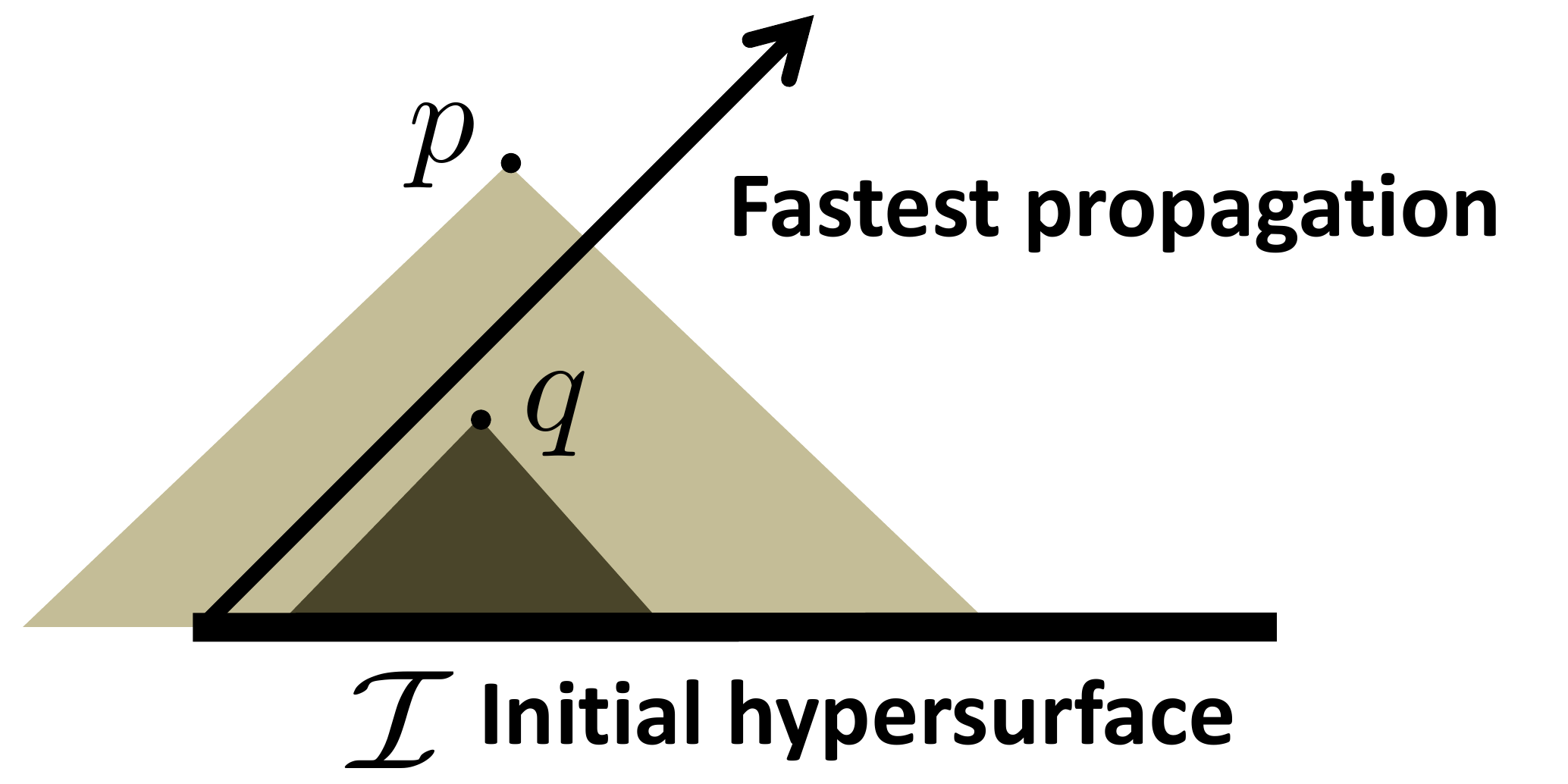}
  \end{center}
  \caption{Relation between the fastest propagation and the edge of Cauchy development: 
  Triangles mean the causal past regions for points $p$ and $q$ based on the fastest propagation, while dotted lines show the light cone from point $p$ defined with null curve. 
} 
    \label{fig1}
\end{figure}

A characteristic hypersurface gives the edge of a Cauchy development, which is related to the \emph{highest propagation speed}. 
The fastest propagation must be tangent to a characteristic hypersurface. 
The intuitive explanation is as follows. 
Suppose that we solve the equation with initial conditions imposed on a hypersurface $\cal I$. (See  Fig.~{\ref{fig1}}.) 
Focusing on point $p$ in Fig.~{\ref{fig1}}, one may say that it is the causal future of hypersurface $\cal I$, if the discussion is based on the \emph{light cone}. 
However, the causal past of $p$ based on all physical propagations including the superluminal modes can reach the outside of initial hypersurface $\cal I$. 
Hence, the physics at $p$ is never uniquely fixed only with the information on $\cal I$ and  
$p$ is located outside of the Cauchy development of $\cal I$. 
Meanwhile, the complete initial conditions on $\cal I$ fix the physics on $q$ uniquely, and thus, $q$ is in the Cauchy development of $\cal I$.
The boundary of the Cauchy development must be described with the fastest propagation. 
Since a characteristic hypersurface shows a boundary beyond which a dynamical equation cannot  be uniquely solved, it is the edge of a Cauchy development. 
Therefore, on a characteristic hypersurface the fastest propagation must propagate.


\section{Gauss-Bonnet Gravity}\label{GB}

We consider  Gauss-Bonnet gravity in a $D$-dimensional spacetime, where the  action is given by
\begin{eqnarray}
S=\int d^Dx\sqrt{-g}\left[ \frac1{2\kappa^{D-2}}\left\{R-2\Lambda+\alpha\left(R^2-4R_{AB}
  R^{AB}+R_{ABCD}R^{ABCD}\right)\right\} + \mathcal{L}_m\right]\label{action},
\end{eqnarray}
where $R_{ABCD}$, $R_{AB}$, $R$ and $\Lambda$ are the Riemann tensor, the Ricci tensor, the Ricci scalar and the cosmological constant, respectively.
$\mathcal{L}_m$ is the Lagrangian for matter fields.
The Gauss-Bonnet parameter is denoted by $\alpha$, which has the dimension of length squared.
We consider the case for $\alpha \ge 0$ where the Einstein vacuum is stable\cite{Boulware:1985wk,Zwiebach:1985uq,Zumino:1985dp,Cho:2001su}.
Since for $D\le 4$ the Gauss-Bonnet term becomes trivial, in this paper cases with $D>4$ are considered.

The equation of motion can be derived from the variation with respect to $g_{AB}$ as
\begin{equation}
G^{AB}+\Lambda g^{AB}-\frac{\alpha}2 \mathcal{H}^{AB}= 2\kappa^{D-2}T^{AB},\label{fieldeq}
\end{equation}
where $G_{AB}$ is the Einstein tensor and  $\mathcal{H}_{AB}$ reads
\begin{eqnarray}
\mathcal{H}_{AB}:=\left(R^2-4R_{CD}R^{CD}+R_{CDEF}
R^{CDEF}\right)g_{AB}-4\left(RR_{AB}-2R_{AC}R_{B}\,^{C}-2R_{ACBD}R^{CD}+R_{ACDE}R_B\,^{CDE}\right).
\nonumber\\
\end{eqnarray}
$T^{AB}$ is the energy-momentum tensor for matter fields. 
We assume that $T^{AB}$ does not include the highest-order derivative of the metric, and thus, it never affects  the characteristics of gravitons.

\section{Characteristics} \label{cha}

We derive the characteristic equations of Gauss-Bonnet gravity. 
The characteristics give the information of the propagating modes. 
In theories with constraints, the structures of the characteristics are generically complicated.
The first-order formalism makes the structure simpler, and thus, in accordance with the technique in Refs.~\cite{Izumi:2013poa,Deser:2013eua,Deser:2013qza} we develop the first-order formalism. 
Then, after reviewing the characteristics in general relativity, we derive the characteristics in  Gauss-Bonnet gravity~\cite{Aragone,ChoquetBruhat:1988dw}.

In the discussion of characteristics, we consider the evolution from a hypersurface. 
We denote the hypersurface by $\Sigma$. 
We define a vector $\xi^A\left(\frac{\partial}{\partial x^A}\right):=\left(\frac{\partial}{\partial t}\right)$ such that $\xi^A$ is not tangent to $\Sigma$. 
We also define a dual vector $\zeta_A dx^A :=dt$. 
Using $\xi^A$ and $\zeta_A$, we can decompose spacetime into the hypersurface $\Sigma$ and the independent direction $\xi^A$ by the projection operator $\top^A_B := \delta^A_B - \xi^A\zeta_B$.
We will denote its action on tensors by Greek indices, i.e.
\begin{eqnarray}
V_\mu:=\top_\mu^A V_A \qquad \mbox{and} \qquad V^\mu:=\top_A^\mu V^A.
\end{eqnarray}
Meanwhile, we denote the contraction of $\xi^A$ and $\zeta_A$ on an index of any tensor by a subscript ``0"  and a superscript ``0" respectively, i.e. $V_0:=\xi^A V_A $ and $V^0:=\zeta_A V^A$.

\subsection{First-order analysis}

The equation of motion is written with the Riemann curvature, only which includes the second-order derivative of the metric. 
We rewrite it in a first-order differential equation with the Levi-Civita connection: 
\begin{eqnarray}
\Gamma_{ABC}:=g_{AD} \Gamma^D_{BC}
=\frac{1}{2}\left(\partial_C g_{AB}+\partial_B g_{AC}
-\partial_A g_{BC}\right). 
\label{Levi}
\end{eqnarray}
This obviously satisfies the symmetric condition of $\Gamma_{ABC}$ with respect to $B$ and $C$. 
The evolution of the metric from the hypersurface $\Sigma$ is obtained from the above definition of the Levi-Civita connection:
\begin{eqnarray}
&&\Gamma_{000}=\frac{1}{2}\partial_0 g_{00} ,\\
&&\Gamma_{\alpha00}=
\frac{1}{2}\left(2\partial_0 g_{0\alpha}-\partial_\alpha g_{00}\right), \\
&&\Gamma_{0\alpha\beta}=\frac{1}{2}\left(\partial_\alpha g_{0\beta}+\partial_\beta g_{0\alpha}
-\partial_0 g_{\alpha\beta}\right) . \label{0ab}
\end{eqnarray}
With the Levi-Civita connection, $R_{\alpha000}$ must be zero, which gives 
\begin{eqnarray}
\partial_0 \Gamma_{00\alpha}-\partial_{\alpha}\Gamma_{000}= f_\alpha [g_{AB},\Gamma_{CDE}].\label{00alpha}
\end{eqnarray}
Here, $f_\alpha [g_{AB},\Gamma_{CDE}]$ is a function of $g_{AB}$ and $\Gamma_{CDE}$. 
Moreover, the Riemann curvature constructed by the Levi-Civita connection satisfies $R_{\beta\gamma0\alpha}-R_{0\alpha\beta\gamma}=0$, which can be written as 
\begin{eqnarray}
\partial_0 \Gamma_{\beta\gamma\alpha}-\partial_\alpha\Gamma_{\beta\gamma0}-\partial_\beta\Gamma_{0\alpha\gamma}+\partial_\gamma \Gamma_{0\alpha\beta} = f_{\alpha\beta\gamma} [g_{AB},\Gamma_{CDE}], \label{alphabetagamma}
\end{eqnarray}
where $f_{\alpha\beta\gamma} [g_{AB},\Gamma_{CDE}]$ is a function of $g_{AB}$ and $\Gamma_{CDE}$. 
Equations (\ref{00alpha}) and (\ref{alphabetagamma}) fix the evolutions of $\Gamma_{00\alpha}$ and $\Gamma_{\alpha\beta\gamma}$. 
Moreover, using Eq.(\ref{0ab}), $\Gamma_{\alpha\beta0}$ can be written as 
\begin{eqnarray}
\Gamma_{\alpha\beta0}&=& \frac{1}{2} (\partial_0 g_{\alpha\beta}+\partial_\beta g_{\alpha 0} - \partial_\alpha g_{\beta 0}) \nonumber\\
&=& -\Gamma_{0\alpha\beta}+ \partial_\beta g_{\alpha 0}. \label{Gab0}
\end{eqnarray}
This is a constraint equation, which fixes the value of $\Gamma_{\alpha\beta0}$\footnote{
K.I. would like to thank James Nester for pointing out the absence of the discussion about $\Gamma_{\alpha\beta0}$.
}.

Now, the time evolutions of $(D+1)D/2$ variables $\Gamma_{000}$, $\Gamma_{\alpha00}$ and $\Gamma_{0\alpha\beta}$ are not fixed yet. 
The gravitational equation of motion fixes $D(D-1)/2$ of them generically, which are physical degrees of freedom.
Meanwhile, the other $D$ degrees of freedom cannot be fixed. 
which are related to the gauge degrees of freedom.
We can easily find that $\partial_0\Gamma_{000}$ and $\partial_0\Gamma_{\alpha00}$ never appear in the form of the Riemann curvature. 
They usually are fixed by hand, i.e. by gauge fixing, or just ignored.
The remaining $D(D-1)/2$ components $\Gamma_{0\alpha\beta}$ must be the physical degrees of freedom, and we discuss their characteristics.

\subsection{Characteristic equation} \label{GBgravity}

Now, we discuss the characteristics only for $\Gamma_{0\alpha\beta}$. 
$\partial_0 \Gamma_{0\alpha\beta}$ appears only in $R_{0\alpha0\beta}(=R_{0\beta0\alpha}=-R_{\alpha00\beta}=\cdots)$,\footnote{
In the previous subsection, $\Gamma_{\alpha\beta0}$ was fixed by the constraint equation (\ref{Gab0}) while we discussed the time evolution for the others. 
Thus, the time derivative of $\Gamma_{\alpha\beta0}$ gives that of $\Gamma_{0\alpha\beta}$ through the constraint equation (\ref{Gab0}), i.e. $\partial_0 \Gamma_{\alpha\beta0} = -\partial_0 \Gamma_{0\alpha\beta}+ \cdots$, 
which joins in the characteristic equation for $\Gamma_{0\alpha\beta}$.
This effect is included in our analysis by considering $R_{\alpha00\beta}$, etc.
}, and thus, we need only to check its coefficient. 
We review characteristics in general relativity and then derive them in Gauss-Bonnet gravity~\cite{Aragone,ChoquetBruhat:1988dw}. 
 
\subsubsection{General relativity} \label{gene}

Terms including $R_{0\alpha0\beta}$ in the Einstein tensor $G^{AB}$  are written as 
\begin{eqnarray}
&&G^{AB}=R_{0\alpha0\beta}A^{AB,\alpha\beta} + \mbox{(other terms)}, \\
&&A^{AB,\alpha\beta}:=g^{\alpha\beta}g^{0A}g^{0B}+g^{00}g^{\alpha A}g^{\beta B}-g^{0\alpha}g^{0A}g^{\beta B}-g^{0\alpha}g^{\beta A}g^{0B}-g^{00}g^{\alpha\beta}g^{AB}+g^{0\alpha}g^{0\beta}g^{AB} .
\end{eqnarray}
Since we can easily check that $A^{AB,\alpha\beta}$ becomes zero for $(A,B)=(0,0),(0,\mu),(\mu,0)$, 
only $(\mu,\nu)$ components are related to characteristic equations. 
$A^{\mu\nu,\alpha\beta}$ can be written simply as
\begin{eqnarray}
A^{\mu\nu,\alpha\beta}=g^{00}(h^{\alpha\mu} h^{\beta\nu}-h^{\alpha\beta}h^{\mu\nu}), \label{Gmncha}
\end{eqnarray}
where $h^{\mu\nu}$ is the inverse matrix of the induced metric on the hypersurface $\Sigma$ and written in terms of $g^{AB}$ as
\begin{eqnarray}
h^{\mu\nu}=g^{\mu\nu}-\frac{g^{0\mu}g^{0\nu}}{g^{00}}.
\end{eqnarray}
Although the form of $h^{\mu\nu}$ seems to give a singular behavior for $g^{00}=0$ where the hypersurface $\Sigma$ becomes null, the singular parts are canceled with each other in Eq.(\ref{Gmncha}). 
Therefore, even if we take the limit as $\Sigma$ approaches to a null hypersurface with 
the above expression, 
$A^{\mu\nu,\alpha\beta}$ is still regular. 
Thus, the limit is continuous to the case on the exact null hypersurface.

We confirm here that the characteristic hypersurface in general relativity becomes null. 
The characteristic equations on the hypersurface $\Sigma$ are 
\begin{eqnarray}
A^{\mu\nu,\alpha\beta}\bar \Gamma_{0\alpha\beta}=0,\label{charaGR}
\end{eqnarray}
where $\bar \Gamma_{0\alpha\beta}$ means it is not the value of the vector $\Gamma_{0\alpha\beta} $, but represents the change of $\Gamma_{0\alpha\beta}$ in a certain direction.
The condition for characteristics is written as
\begin{eqnarray}
0&=&\det \left[ A^{\mu\nu,\alpha\beta} \right] \nonumber\\
&=&(-1)^{D-1}(D-2) (\det [g_{AB}])^{-\frac{D(D-1)}{2}}(\det [h^{\mu\nu}])^{-\frac{D(D-3)}{2}},
\end{eqnarray}
where, taking the determinant $\det \left[ A^{\mu\nu,\alpha\beta} \right] $ on the first line,  we consider two combinations $(\mu,\nu)$ and $(\alpha,\beta)$ as two indices of the rank-2 matrix.
While $\det [g_{AB}]^{-1}$ must be nonzero on a regular manifold, $\det [h^{\mu\nu}]^{-1}$ can be zero if and only if the hypersurface $\Sigma$ is null. 
Therefore, the characteristic hypersurfaces for gravitons in general relativity are always null.

For our later discussion, we  check which components of the equation become characteristics. 
Since now we know that the characteristic hypersurface is null, we consider a null hypersurface.
We can always diagonalize and normalize the induced metric at a point as 
\begin{eqnarray}
\left(h^{-1}\right)_{\mu\nu}=\mbox{diag} (0,1,1,\ldots ,1). \label{diagh}
\end{eqnarray}
We use the index ``$1$" for the first component, i.e. the null direction, while the others are labeled with $(i,j,\ldots )$.
$h^{\mu\nu}$ diverges as $O[(g^{00})^{-1}]$ only for $(\mu,\nu)=(1,1)$, while the others are finite, which are $h^{1i}=0$ and $h^{ij}=\mbox{diag}(1,1,\ldots ,1)$. 
Because $A^{\mu\nu,\alpha\beta}$ is proportional to $g^{00}$, which is zero on a null hypersurface, without $h^{11}$ it becomes zero. 
Then, Eq.(\ref{charaGR}) becomes
\begin{eqnarray}
-g^{00}h^{11}\sum_i  \bar\Gamma_{0ii}&=&0 \qquad [\mbox{for }(1,1)-\mbox{component}] , \label{chara11}\\
\frac{1}{2}g^{00}h^{11}  \bar\Gamma_{01i}&=&0 \qquad [\mbox{for }(1,i)-\mbox{component}] ,\label{chara1i}\\
-g^{00}h^{11} \bar\Gamma_{011} \delta_{ij}&=&0 \qquad [\mbox{for }(i,j)-\mbox{component}] .\label{charaij}
\end{eqnarray}
Only $\bar\Gamma_{011}$ appears in Eq.(\ref{charaij}), and thus, there are $D(D-3)/2$ degeneracies.
Equations (\ref{chara11}) and (\ref{chara1i}) fix $\bar\Gamma_{01i}$ and the trace of $\bar\Gamma_{0ij}$.
As a result, we cannot fix totally $D(D-3)/2$ of $\bar\Gamma_{0ij}$, which are traceless components of $\bar\Gamma_{0ij}$. 
The number of degrees of freedom is equal to that of gravitational propagations.
Since $\Gamma_{0ij}$ includes $\partial_0 g_{ij}$, these characteristics are related to the propagations of traceless components of $g_{ij}$. 
Moreover, the null direction labeled with $1$ is transverse direction for $g_{ij}$. 
Therefore, these characteristics are corresponding to all of gravitational modes propagating in $1$ direction.

\subsubsection{Gauss-Bonnet gravity}\label{gb}
The terms including $R_{0\alpha0\beta}$ in $\mathcal{H}^{AB}$ are written in
\begin{eqnarray}
&&\mathcal{H}^{AB} = R_{0\alpha0\beta}B^{AB,\alpha\beta} + \mbox{(other terms)},\\
&&B^{AB,\alpha\beta} :=(4g^{00}g^{\alpha\beta}g^{AB}R-4g^{0\alpha}g^{0\beta}g^{AB}R-8g^{00}g^{AB}R^{\alpha\beta}-8g^{\alpha\beta}g^{AB}R^{00}+16g^{0\alpha}g^{AB}R^{0\beta} \nonumber\\
&&\qquad\qquad\qquad
-8g^{00}g^{\alpha\beta}R^{AB}+8g^{0\alpha}g^{0\beta}R^{AB}-4g^{00}g^{\alpha A}g^{\beta B}R+4g^{0\alpha}g^{0A}g^{\beta B}R+4g^{0\alpha}g^{0B}g^{\beta A}R\nonumber\\
&&\qquad\qquad\qquad
-4g^{\alpha\beta}g^{0A}g^{0B}R+8g^{00}g^{\alpha A}R^{\beta B}+8 g^{00}g^{\alpha B}R^{\beta B}+8g^{\alpha\beta}g^{0A}R^{0B}+8g^{\alpha\beta}g^{0B}R^{0A}\nonumber\\
&&\qquad\qquad\qquad
-8g^{0\alpha}g^{0A}R^{\beta B}-8 g^{0\alpha}g^{0B}R^{\beta A}
-8g^{0\alpha}g^{\beta A}R^{0B} -8 g^{0\alpha}g^{\beta B}R^{0A}+8g^{0A}g^{0B}R^{\alpha\beta}\nonumber\\
&&\qquad\qquad\qquad
+8g^{\alpha A}g^{\beta B}R^{00}-8g^{\alpha A}g^{0B}R^{0\beta}-8g^{\alpha B}g^{0A}R^{0\beta}+8g^{AB}R^{0\alpha0\beta}+8g^{\alpha\beta}R^{0A0B}+8g^{00}R^{\alpha A \beta B}\nonumber\\
&&\qquad\qquad\qquad
-8g^{0\alpha}R^{A\beta B 0}-8 g^{0\alpha}R^{B\beta A0}-8g^{0A}R^{B\alpha 0\beta}-8g^{0B}R^{A\alpha 0\beta}-8 g^{\alpha A}R^{B0\beta 0}-8g^{\alpha B}R^{A0 \beta0}).
\end{eqnarray}
For $(A,B)=(0,0),(0,\mu),(\mu,0)$, $B^{AB,\alpha\beta}$ gives zero. 
The other components become
\begin{eqnarray}
&&B^{\mu\nu,\alpha\beta}= 4g^{00}R_{\lambda\omega\gamma\delta}(h^{\lambda\gamma}h^{\omega\delta}h^{\mu\nu}h^{\alpha\beta}-h^{\lambda\gamma}h^{\omega\delta}h^{\mu\alpha}h^{\nu\beta}
+2h^{\lambda\mu}h^{\gamma\alpha}h^{\omega\delta}h^{\nu\beta}+2h^{\lambda\nu}h^{\gamma\alpha}h^{\omega\delta}h^{\mu\beta}\nonumber\\
&&\qquad\qquad\qquad\qquad\qquad\qquad\qquad
-2h^{\lambda\alpha}h^{\gamma\beta}h^{\omega\delta}h^{\mu\nu}-2h^{\lambda\mu}h^{\gamma\nu}h^{\omega\delta}h^{\alpha\beta}+2h^{\lambda\mu}h^{\omega\alpha}h^{\gamma\nu}h^{\delta\beta}).
\end{eqnarray}
Although this also looks singular in the case where the hypersurface $\Sigma$ is null, 
the singular parts are canceled out and it becomes finite. 
Note that $B^{\mu\nu,\alpha\beta}$ does not involve $R_{0\alpha0\beta}$. 
Therefore, the equation of motion does not include the square of $\partial_0 \Gamma_{0\alpha\beta}$.
It is a notable property of Gauss-Bonnet gravity, which makes the time evolution unique. 

The characteristic equations for gravitons in Gauss-Bonnet gravity are composed of the Einstein-Hilbert and the Gauss-Bonnet components:
\begin{eqnarray}
\left(A^{\mu\nu,\alpha\beta}-\frac{\alpha}{2}B^{\mu\nu,\alpha\beta} \right)\bar \Gamma_{0\alpha\beta}=0.
\end{eqnarray} 
Let's see that the characteristic hypersurface is generically not null.
We see how eqs.(\ref{chara11}-\ref{charaij}) are modified.
Each component is written with a diagonalized and normalized induced metric (\ref{diagh}) as
\begin{eqnarray}
-g^{00}h^{11}\left[\sum_i  \bar\Gamma_{0ii}+2\alpha\left( \sum_{i,k,l}R_{klkl} \bar\Gamma_{0ii} -2 \sum_{i,j,k}R_{ikjk} \bar\Gamma_{0ij} \right)\right]=0  \qquad\qquad\qquad\qquad [\mbox{for }(1,1)-\mbox{component}]  ,\qquad \label{GBchara11}&&  \\
\frac{1}{2}g^{00}h^{11} \left[ \bar\Gamma_{01i} +2\alpha\left( \sum_{k,l}R_{klkl} \bar\Gamma_{01i} -2 \sum_{j,k}R_{ikjk} \bar\Gamma_{01j} \right)  +8\alpha\sum_{j,k}\left(R_{1kik}\bar\Gamma_{0jj}-R_{1kjk}\bar\Gamma_{0ij}-R_{1jik}\bar\Gamma_{0jk}\right) \right]=0 \quad && \nonumber\\
\qquad [\mbox{for }(1,i)-\mbox{component}]  ,\qquad \label{GBchara1i}&& \\
-g^{00}h^{11} \left[\delta_{ij}\bar\Gamma_{011} +2\alpha\left(\sum_{k,l}R_{klkl}\delta_{ij}-2R_{ikjk} \right) \bar\Gamma_{011} +\alpha\sum_k\left( R_{1ijk}+R_{1jik} \right)\bar\Gamma_{01k}\right. \qquad\qquad\qquad\qquad\qquad\qquad &&\nonumber\\
\left.
+4\alpha\left\{\delta_{ij}\sum_{k,l}\left(R_{1k1k}\bar\Gamma_{0ll}-R_{1k1l}\bar\Gamma_{0kl}\right)
+\sum_k\left(R_{1i1k}\bar\Gamma_{0kj}+R_{1j1k}\bar\Gamma_{0ki}-R_{1k1k}\bar\Gamma_{0ij}-R_{1i1j}\bar\Gamma_{0kk}\right)\right\}\right] 
 =0&&\nonumber\\
  \qquad [\mbox{for }(i,j)-\mbox{component}]  .\qquad&& \label{GBcharaij}
\end{eqnarray}
We can find that the components of Eq.(\ref{charaij}), which are degenerated on a null hypersurface in  general relativity, are modified and with a generic form of $R_{ABCD}$ the degeneracies are resolved. 
We can easily see it by considering a simple example, where $R_{ijkl}=0$, $R_{1ijk}=0$ and $R_{1i1j}= C \delta_{ij}$. 
Then, Eqs.(\ref{GBchara11}) and (\ref{GBchara1i}) become the same as those in general relativity, i.e. Eqs.(\ref{chara11}) and (\ref{chara1i}), while the 
$(i,j)$ component (\ref{GBcharaij}) becomes 
\begin{eqnarray}
-g^{00}h^{11}\left[\delta_{ij}\bar\Gamma_{011}+4\alpha(D-4)C\left(\delta_{ij} \sum_k \bar\Gamma_{0kk}-\bar\Gamma_{0ij}\right)\right]=0.
\end{eqnarray}
In the above equation, we can see that the degeneracies are completely resolved.
Therefore, this null hypersurface is not characteristic. 

For $R_{1ijk}=0$ and $R_{1k1l}=0$, in contrast, the structure of characteristic equation is the same as that in general relativity. 
Namely, all components of Eq.(\ref{GBcharaij}) are equations for $\bar \Gamma_{011}$ or trivial, and thus, they are still degenerate.
$D(D-3)/2$ degrees of freedom of $\bar \Gamma_{0ij}$ eventually cannot be fixed as in  general relativity.
This means that the null hypersurface is still characteristic.

\section{Causal structures}\label{cau}

Now we have the characteristic equation of  Gauss-Bonnet gravity. 
Using it, we can analyze the causal structures. 
Firstly, we consider stationary solutions and find that Killing horizons express the causal edges, 
i.e. a black hole horizon in the sense of causality. 
Secondly, we consider $(D-2)$-dimensionally maximally symmetric solutions without the stationary assumption. 
We show that, if the geometrical null energy condition is satisfied, the speeds of gravitons must be less than or equal to that of light. 
On the other hand, on evaporating black holes where the geometrical null energy condition is expected to be broken, gravitons can propagate faster than light.

\subsection{Locally stationary cases} \label{LS}
In Sec.\ref{GBgravity}, we saw that if on a null hypersurface the conditions $R_{1ijk}=0$ and $R_{1i1j}=0$ are satisfied, the hypersurface is characteristic for all degrees of freedom of gravitons. 
We shall see a sufficient condition for this. 

Here, we consider the case where the hypersurface  $\Sigma$ is null. 
We denote the direction normal to $\Sigma$ by the label $1$;
i.e. with the normal null vector $n^A$ we have $V_1:=n^A V_A$. 
The normal vector $n^A$ lies on the hypersurface $\Sigma$. 
The Latin indices $(i,j,\ldots )$ label the other spacelike directions normal to $n^A$ on the hypersurface $\Sigma$. 
Since all of vectors lying on the hypersurface $\Sigma$ must be normal to $n^A$, 
we have $g_{11}=0$ and $g_{1i}=0$ on the hypersurface $\Sigma$. 
Therefore, their higher-order derivatives with respect to $\partial_1$ and $\partial_i$, i.e. $\partial_\mu\ldots \partial_\nu g_{11}$ and $\partial_\mu\ldots \partial_\nu g_{1i}$, must be zero. 
Furthermore, the conditions $g_{11}=0$ and $g_{1i}=0$ lead to $g^{00}=0$ and $g^{0i}=0$. 
Imposing the additional conditions 
\begin{eqnarray}
\partial_1 g_{ij}=0, \qquad \partial_1^2g_{ij}=0 \qquad \mbox{and} \qquad\partial_1\partial_k g_{ij}=0, \label{sufcon}
\end{eqnarray}
together with the above conditions we can obtain $R_{1ijk}=0$ and $R_{1i1j}=0$ by direct calculation. 
Thus, a combination of Eqs.(\ref{sufcon}) is a sufficient condition for the null hypersurface $\Sigma$ to be characteristic. 

On a Killing horizon, the normal null vector $n^A$ is the Killing vector, which results in $n^A \partial_A g_{\mu\nu}= \partial_1 g_{\mu\nu}=0$.
Combined with the fact that the label ``$k$" is the index for the tangent direction to hypersurface $\Sigma$, we can find that Eqs.(\ref{sufcon}) are always satisfied on Killing horizon.
Therefore, on a stationary solution such that the Killing horizon is coincident with the event horizon defined by null curves, i.e. 
the event horizon is exactly the causal edge for gravitons. 
Classical gravitons never come out from inside of stationary black holes.

\subsection{$(D-2)$-dimensionally maximally symmetric cases} \label{DB}
On a generic spacetime, Eqs.(\ref{sufcon}) are not satisfied. 
Then, it is important to see how the characteristic hypersurface is modified, 
i.e. whether it becomes spacelike or timelike. 
A spacelike characteristic results in the existence of a superluminal mode, 
which breaks the discussion of causal structures based on null curves. 
Here, for simplicity, we consider cases with a maximally symmetric ${D-2}$ dimensional space, where the metric can be generically written as 
\begin{eqnarray}
ds^2= -2f(u,v) du dv +\left[R(u,v)\right]^2 d\Omega_{D-2}^2 \ . \label{met}
\end{eqnarray}
We choose both $U^A:=(\partial/\partial u)^A$ and $V^A:=(\partial/\partial v)^A$ to be future pointing null vectors; i.e. $f(u,v)$ is positive.
$d\Omega_{D-2}^2$ is the ${D-2}$ dimensional metric that is  maximally symmetric, constant and spacelike. 
The metric component for $d\Omega_{D-2}^2$ is defined as 
\begin{eqnarray}
d\Omega_{D-2}^2 := \gamma_{ij} dx^i dx^j.
\end{eqnarray}
$f(u,v)$ and $R(u,v)$ are functions of $u$ and $v$. 
We consider a maximally symmetric ${D-2}$ dimensional hypersurface $\Sigma$, on which $\bar v:=v+\epsilon u$ is constant. 
It is convenient to use new coordinate variables $\bar u:= u$ and $\bar v$, with which the metric (\ref{met}) is written in
\begin{eqnarray}
ds^2= - 2f d\bar u d\bar v+ 2 \epsilon f {d\bar u}^2+R^2 d\Omega_{D-2}^2. \label{metric}
\end{eqnarray}
${\bar U}^A:=(\partial/\partial \bar u)^A$ lies on the hypersurface $\Sigma$ and  
${\bar V}^A:=(\partial/\partial \bar v)^A (=V^A)$ is a null vector that is never tangent to $\Sigma$.
For $\epsilon>0$, $\epsilon<0$ or $\epsilon=0$, the hypersurface $\Sigma$ is spacelike, timelike or null, respectively. 

First of all, we show that, if $R_{AB}U^A U^B =0$, the hypersurface for $v=$const is characteristic. 
Seeing Eq.(\ref{GBcharaij}), we know that for $R_{AiBj}U^A U^B =0$ all degeneracies are never resolved. 
Because of the symmetry, $R_{AiBj}U^A U^B $ must be proportional to $g_{ij}$, i.e. $R_{AiBj}U^A U^B = C g_{ij}$. 
Since the directions labeled with $(i,j,\ldots )$ are normal to two null vector $U^A$ and $V^A$, 
we have
\begin{eqnarray}
R_{AB}U^A U^B = R_{iAjB}U^AU^B g^{ij} =(D-2) C.
\end{eqnarray}
Therefore, $R_{AB}U^A U^B =0$ results in $C=0$, which gives $R_{AiBj}U^A U^B = C g_{ij}=0$,
 and the characteristic hypersurface is null.

Next, we consider cases where $R_{AB}U^A U^B \neq 0$. 
We have $R_{AiBj}U^A U^B = C g_{ij}$ with nonzero $C$ and the sign of $C$ is coincident with that of $R_{AB}U^A U^B$. 
This makes the degeneracies of Eq.(\ref{GBcharaij}) resolved and 
shifts the characteristic hypersurface. 
Since the characteristic hypersurface becomes non-null, i.e. $\epsilon\neq 0$, 
by considering the effect of shifting the hypersurface, we have a modification of Eq.(\ref{GBcharaij}),
which is originally derived on a null hypersurface.
Both the Einstein tensor and the Gauss-Bonnet term give corrections, but
in regions with small curvature\footnote{
Gauss-Bonnet gravity is the low-energy effective theory of Lovelock gravity obtained by ignoring the higher curvature terms called Lovelock terms~\cite{Lovelock:1971yv}.
Smallness of curvature is required for it to be valid to ignore these terms..
} 
the correction arising from the Gauss-Bonnet term is negligibly small compared to that coming from the Einstein tensor. 
Therefore, we ignore the correction stemming from the Gauss-Bonnet term.
We show, by the discussion of the balance between the modifications stemming from the violation of $R_{AB}U^A U^B =0$ and from the non-nullity of hypersurface $\Sigma$, 
whether the hypersurface $\Sigma$ becomes spacelike or timelike. 
To see this, we only need to check the sign of $\epsilon$. 


For $R_{AB}U^A U^B = (D-2) C  \neq 0$, the coefficients of $\bar \Gamma_{0ij}$ in Eq.(\ref{GBcharaij})  read 
\begin{eqnarray}
&&-4 \alpha g^{00} h^{11} \left\{\delta_{ij}\sum_{k,l}\left(R_{1k1k}\bar\Gamma_{0ll}-R_{1k1l}\bar\Gamma_{0kl}\right)
+\sum_k\left(R_{1i1k}\bar\Gamma_{0kj}+R_{1j1k}\bar\Gamma_{0ki}-R_{1k1k}\bar\Gamma_{0ij}-R_{1i1j}\bar\Gamma_{0kk}\right)\right\} \nonumber\\
&&\qquad
= 4 \alpha (D-4)C f^{-2} \left(g^{ij}g^{kl}-g^{ik}g^{jl}\right) \bar \Gamma_{0kl}.
\end{eqnarray}
The contribution of shifting the hypersurface from null can be obtained from Eq.(\ref{Gmncha}) as
\begin{eqnarray}
A^{ij,kl}\bar \Gamma_{0kl} = 2\epsilon f^{-1}\left(g^{ij}g^{kl}-g^{ik}g^{jl}\right) \bar \Gamma_{0kl},
\end{eqnarray}
where, as we commented, we ignore the contribution coming from the Gauss-Bonnet term. 
The degeneracy in the modified equation happens only when these two contributions cancel.
Namely, the condition is 
\begin{eqnarray}
4\alpha  (D-4)C f^{-2}+2\epsilon f^{-1}=0 \qquad\Leftrightarrow \qquad
\epsilon=-2\alpha (D-4) C f^{-1}.
\end{eqnarray}
For $D\le 4$ the Gauss-Bonnet term becomes trivial, and thus $D$ must be larger than four. 
We have $\alpha>0$ for the stability of the Einstein vacuum and set $f$ to be positive. 
As a result, the sign of $\epsilon$ is opposite to that of $C$, i.e. $R_{AB}U^A U^B$. 

If the null energy condition is satisfied in the geometrical sense, that is $R_{AB}N^A N^B\ge 0$  where $N^A$ is any null vector\footnote{In the general relativity through the Einstein equation we can show that the geometrical null energy condition $R_{AB}N^A N^B\ge 0$ is equivalent to the null energy condition defined with energy-momentum tensor $T_{AB}N^A N^B\ge 0$;
however in a general gravity theory we do not have the equivalence. 
Here, ``geometrical sense" means that the condition is the same as that rewritten in geometrical terms with the Einstein equation.
}, $\epsilon$ is always nonpositive, and thus the characteristics is timelike or null. 
This means that the speed of gravitons in the radial direction on a spherically symmetric spacetime and of gravitational plane waves on flat space is less than or equal to that of light.
The equality happens only for  $R_{AB}U^A U^B=0$.
Gravitons do not break the causal structure.\footnote{
If the speeds of all  fields become less than that of light, the causal structure must be modified.
}

On the other hand, if the geometrical null energy condition is violated, we potentially have superluminal modes in the radial direction. 
Namely, for $R_{AB}U^A U^B< 0$ the characteristic becomes spacelike.
There may be two possibilities for it to happen. 
In Gauss-Bonnet gravity, the null energy condition for matter fields $T_{AB}N^A N^B\ge 0$ does not result in the geometrical null energy condition. 
In the case with $(D-2)$-dimensional maximal symmetry, the relation between the null energy condition for matter fields and the geometrical one has been investigated~\cite{Nozawa:2007vq,Maeda:2007uu,Maeda:2011ii}.\footnote{
K.I. would like to thank Hideki Maeda for pointing this out.
}
We have two branches of solutions; the Einstein branch and the Gauss-Bonnet branch. 
The definition of the Einstein branch is a sequence of solutions where the generalized Misner-Sharp quasilocal mass approaches to the original one in the limit as the Gauss-Bonnet parameter $\alpha$ goes to zero.
If not, the solution belongs to the Gauss-Bonnet branch. 
On the Einstein branch the sign of $T_{AB}U^A U^B$ coincides with that of $R_{AB}U^A U^B$, while on the Gauss-Bonnet branch it becomes opposite.
Therefore, on the Gauss-Bonnet branch, if we impose the null energy condition on matter fields, $R_{AB}U^A U^B< 0$ can occur. 
The other possibility stems from the quantum effects on curved space time. 
Considering the backreaction of the Hawking radiation, 
we expect that the black hole is shrinking, i.e. its area is decreasing. 
With a decreasing area of a black hole, it is necessary to break the geometrical null energy condition. 
Exactly stated, we need the violation of $R_{AB}U^A U^B\ge 0$ for the outgoing null vector $U^A$. 
As a result, for evaporating black holes, null hypersurfaces are not the boundary of causally related region. 
In other words, gravitons can escape from ``black holes" defined with null curves.

\section{Summary}\label{Sum} 

We have analyzed causal structures in  Gauss-Bonnet gravity. 
A theory with noncanonical kinetic terms potentially has superluminal propagations. 
Gauss-Bonnet gravity is one such theory. 
This was pointed out at the end of the 1980s~\cite{Aragone,ChoquetBruhat:1988dw}, and that the concrete  solutions having superluminal propagations was shown in Refs.~\cite{Brigante:2007nu,Brigante:2008gz}. 
Superluminal propagations make causal structures complicated. 
We should discuss causal structures based not on null curves but on the fastest modes. 

To analyze causal structures, we derived the characteristic equations in Sec.~\ref{cha}. 
There, we have \emph{not} fixed the gauge degrees of freedom, and obtained a result consistent with Refs.~\cite{Aragone,ChoquetBruhat:1988dw}.
In our formalism,  since the vector $\xi^A$ is not needed to be normal to the hypersurface $\Sigma$ unlike in Refs.~\cite{Aragone,ChoquetBruhat:1988dw}, we can take the smooth limit as $\Sigma$ approaches to a null hypersurface.
We have demonstrated that with a generic curvature the Gauss-Bonnet effect resolves the degeneracies of characteristic equations on null hypersurface.
The resolution means that the directions of the graviton propagations are not null. 

With the characteristic equations, we have analyzed the causal structures in Sec.~\ref{cau}. 
We have discussed locally stationary cases in Sec~\ref{LS}. 
We have proved that if Eqs.(\ref{sufcon}) are satisfied on a null hypersurface, 
the hypersurface is a characteristic. 
Since on Killing horizons these conditions are always satisfied, they are exactly the edges of Cauchy development for gravitons. 
Namely, on stationary spacetime, we can trust that the causal edges based on null curves and Killing horizons become exactly the event horizons in the sense of causality. 

We have also analyzed the causal structure on solutions with ${D-2}$ dimensional maximal symmetry  in Sec.~\ref{DB}. 
We have shown that, if the null energy condition in the geometrical sense is satisfied, 
the radial velocities of gravitons are less than or equal to that of light. 
On the Einstein branch, the geometrical null energy condition holds if the null energy condition on matter fields does. 
Since the existence of subluminal modes does not change the fact that photons have the maximum speed in the theory, it does not break the discussions of causal structures based on null curves.
Namely,  nothing harmful appears in the sense of causality. 
However, with subluminal gravitons gravitational Cherenkov radiation may occur.
Since the Gauss-Bonnet term appears only in higher-dimensional theory, 
we need to consider the compactification of higher dimensions~\cite{Giribet:2006ec} or the braneworld models~\cite{Kim:1999dq,Cho:2001nf,deRham:2006pe,Kofinas:2003rz,Brown:2006mh,Charmousis:2002rc,Maeda:2003vq,Maeda:2007cb,BouhmadiLopez:2012uf,Bouhmadi-Lopez:2013gqa,Yamashita:2014cra,Liu:2014xja}.
They could be constrained from observational results~\cite{Moore:2001bv,Kimura:2011qn}.\footnote{K.I. would like to thank Takahiro Tanaka for pointing this out and Shinji Mukohyama for suggesting the concrete methods.}

On the other hand, if the geometrical null energy condition is violated, the radial propagation of gravitons on a spherically symmetric space can be faster than light. 
Exactly stated, if $R_{AB} U^A U^B<0$ is satisfied for an outgoing (or ingoing) null vector $U^A$, 
the outgoing (ingoing) propagations of gravitons are faster than light. 
It happens on the Gauss-Bonnet branch if we impose the null energy condition on matters, 
i.e. $T_{AB} U^A U^B\ge 0$. 
Moreover, if we consider the backreaction of the Hawking radiation to gravity, 
the geometrical null energy condition should be violated. 
Considering the backreaction of the Hawking radiation, particles with energy are emitted, and thus, the mass of the black hole is decreasing. 
This leads to the decreasing of its area. 
In contrast, with the geometrical null energy condition, the area of a black hole must increase. 
Consequently, on an evaporating black hole, we have a violation of the geometrical null energy condition. 

Considering some UV-complete theories, such as  superstring theory, 
noncanonical kinetic terms appear in their effective theory. 
As we have demonstrated in the case of Gauss-Bonnet gravity, such a theory potentially has superluminal modes. 
Superluminal modes may be one of the solutions for the information loss paradox of an evaporating black hole. 
When we discuss an evaporating black hole, we use the semiclassical approach of gravity.
Namely, we quantize matter fields, while gravity is classical. 
It is possibly hard to discuss the causal structure of a quantum system.
Using graviton propagation, we do not bother with the difficulty of quantum systems. 
Therefore, Gauss-Bonnet gravity is a good objective to see the effect stemming from the noncanonical kinetic terms on an evaporating black hole. 
We expect that quantized matter fields have similar causal structures.
We have the result that on an evaporating black hole the propagation of gravitons is faster than light. 
This means that classical gravitational waves can escape from a black hole that is defined by null curves. 
Therefore, the event horizon defined with null curves is not the edge of the causal region and 
information can easily leak from this  event horizon. 
In the region where curvature becomes large, the higher curvature terms of Lovelock theory~\cite{Lovelock:1971yv} become more dominant and the causal structures are expected to be more nontrivial.

Conversely, superluminal modes might be prohibited by the discussion of UV completion~\cite{Adams:2006sv}.
If so, a large $\alpha$ is forbidden because the solution with superluminal gravitons was found in Refs.~\cite{Brigante:2007nu,Brigante:2008gz} for $\alpha> -\frac{\Lambda}{400}$ with a negative cosmological constant $\Lambda$ in a five-dimensional spacetime. 
Considering more general solutions in Gauss-Bonnet gravity, 
the superluminal modes
could easily appear 
because there is no direct way to confine the geometrical energy condition from the energy condition on matter fields through the equation of motion. 
This may result in the end of the theory.
Therefore, if the existence of superluminal modes is prohibited, 
we need some mechanism to remove out solutions with the superluminal modes such as the Gauss-Bonnet branch, for instance, by 
revising the gravitational equation with nontrivial forms of matter actions or its coupling with gravity.
The problem might be related to the nonlinear quantum instability of the Einstein vacuums in Gauss-Bonnet gravity~\cite{Charmousis:2008ce}.

In this paper, we have concentrated on Gauss-Bonnet gravity, which is the lowest-order correction of Lovelock gravity. 
We expect the same property to hold in  Lovelock gravity and will analyze it in the future. 
Without the stationary assumption, we have analyzed only the metric with ${D-2}$ dimensional maximal symmetry, but it would be interesting to discuss more general cases, which are also our future works.

\section*{Acknowlegement}

Keisuke Izumi would like to thank Pisin Chen, Shou-Huang Dai, Je-An Gu, Lance Labun, Yen-Wei Liu, Hideki Maeda, Shinji Mukohyama, Takashi Nakamura, James Nester, Yen Chin Ong, Ryo Saito, Testuya Shiromizu, Tadashi Takayanagi, Takahiro Tanaka and Wen-Yu Wen for valuable discussions and helpful comments. 
Keisuke Izumi is supported by Taiwan National Science Council under Project No. NSC101-2811-M-002-103.


\end{document}